\documentclass{ifacconf}

\usepackage{graphicx}      
\usepackage{natbib}        
\usepackage{amsmath,amssymb,amsfonts}

\usepackage{graphicx}
\usepackage{textcomp}
\usepackage{tabu}
\usepackage{algpseudocode}
\usepackage{booktabs}
\usepackage{multirow}
\usepackage{booktabs}
\newtheorem{Remark}{Remark}

\usepackage{epstopdf}
\usepackage{pbox}
\usepackage{algorithm}
\usepackage{CJK}
\usepackage{mathtools}%

 \usepackage{color}

\usepackage[short]{optidef}
 \usepackage{flushend}

\begin{document}
\begin{frontmatter}

\title{Online Ecological Gearshift Strategy via Neural Network with Soft-Argmax Operator\thanksref{footnoteinfo}} 
\thanks[footnoteinfo]{This work is supported by the International Technology Cooperation Program of Science and Technology Commission of Shanghai Municipality (No.21160710600), the National Nature Science Foundation of China (No.62003244), and the Fundamental Research Funds for the Central Universities.}

\author[First]{Xi Luo} 
\author[Second]{Shiying Dong} 
\author[Third]{Jinlong Hong\thanksref{footnoteauthor}}
\thanks[footnoteauthor]{Corresponding author.}
\author[Fourth]{Bingzhao Gao}
\author[Fifth]{Hong Chen}

\address[First]{School of Automotive Studies, Tongji University, Shanghai, China, (e-mail: luoxi21@tongji.edu.cn)}
\address[Second]{Department of Control Science and Engineering, Jilin University, Changchun, China. (e-mail: , (email: sydong19@mails.jlu.edu.cn)}
\address[Third]{School of Automotive Studies, Tongji University, Shanghai, China, (e-mail: hongjl@tongji.edu.cn)}
\address[Fourth]{New Energy Vehicle Engineering Center, Tongji University, Shanghai, China, (e-mail: gaobz@tongji.edu.cn)}
\address[Fifth]{College of Electronic and Information Engineering, Tongji University, Shanghai, China, (email: chenhong2019@tongji.edu.cn)}

\begin{abstract}                
This paper presents a neural network optimizer with soft-argmax operator to achieve an ecological gearshift strategy in real-time. 
The strategy is reformulated as the mixed-integer model predictive control (MIMPC) problem to minimize energy consumption. Then the outer convexification is introduced to transform integer variables into relaxed binary controls. 
To approximate binary solutions properly within training, the soft-argmax operator is applied to the neural network with the fact that all the operations of this scheme are differentiable. 
Moreover, this operator can help push the relaxed binary variables close to 0 or 1. 
To evaluate the strategy effect, we deployed it to a 2-speed electric vehicle (EV). 
In contrast to the mature solver \texttt{Bonmin}, our proposed method not only achieves similar energy-saving effects but also significantly reduces the solution time to meet real-time requirements. This results in a notable energy savings of 6.02\% compared to the rule-based method.
\end{abstract}


\end{frontmatter}

\section{Introduction}
The increasing energy consumption and environmental issues pose formidable challenges to energy efficiency and emissions reduction in vehicles. Meanwhile, during daily driving conditions, numerous factors, such as powertrain working conditions, dynamic traffic, and road geometry could significantly influence the energy consumption characteristics (\cite{dong2021hierarchical}). Advanced powertrain control is considered to be an effective step to achieve high efficiency and low emission.
\par
Generally, the widespread presence of transmissions in automobiles poses an optimization problem for gear selection during driving operations, which will affect the working point of engines or motors. In this regard, several concepts for ecological gearshift have been proposed. The gearshift schedule is mainly divided into two categories, namely rule-based method and optimization-based method. As for rule-based method, the look-up table is mostly used to reduce the online calculation burden to the greatest extent, which is usually designed by off-line analysis~(\cite{ngo2014gear}) or global optimization~(\cite{xu2019formulation}), e.g. dynamic programming (DP), iterative search. And it won't take the foreseeing working condition into account that using off-line analysis, leading to instant optimization. Besides, DP is usually designed under specific driving cycles, lacking application generality.
\par
The gearshift schedule is usually reformulated as a mixed-integer model predictive control problem. While travelling, the speed profile and gearshift are both needed to be optimized for energy consumption. Linear programming solver is applied for speed profile and then DP is introduced to achieve gear position~(\cite{sundstrom2019optimal}). Recently, machine learning empowers optimization control solutions and reinforcement learning is used for traction force and gear shift determination~(\cite{li2019ecological}). To achieve the integer controls, the softmax activation is applied for shift decisions based on its capability for classification. However, mixed-integer programming (MIP) for gearshift decisions still poses the challenge for real-time solutions in the vehicle chip controller. In order to get the controls fast enough, learning-based methods are proposed. The supervised learning based on large amounts of labelled data is invested for integer inputs~(\cite{karg2018deep}). Apart from this, some researches have applied the characteristics of neural networks to imitate the traditional MIP solution to help accelerate the solving speed, such as branch-and-bound~(\cite{gasse2019exact}), large neighborhood search method~(\cite{sonnerat2021learning}). But the aforementioned methods mostly rely on the structure of the MIP problem and still spend too much effort to seek an exact integer solution.
\par
Drawing inspiration from the solution methods for continuous optimization problems in~\cite{drgona2020learning}, which utilizes the automatic differentiation of neural network training process to achieve model predictive control (MPC) open-loop optimization. We first relax the gear integer controls into continuous variables and formulate an MIMPC problem for the ecological gearshift. Then the neural network is applied to achieve the open-loop optimization, in which the soft-argmax is to approximate the integer controls. As for the closed-loop control, the necessary operation, such as the standard rounding strategy is applied for strict integer controls.
\par
The main contributions are: (1) A neural network optimizer scheme with soft-argmax operator is proposed for MIMPC to approximate binary controls, which is differentiable for updating parameters. (2)The method we proposed has yielded satisfactory results in both solution quality and solution speed compared to the mature MIP solver, e.g., \texttt{Bonmin}, which indicates to be deployed online.
\par
The rest of this paper is organized as follows: Section \ref{sec: Model} builds the electric vehicle dynamics, and formulates the MIMPC problem, including the outer convexification. The neural network optimizer is presented with soft-argmax operator, the parameter updating method is derived, and the closed loop application is proposed in section \ref{sec: NN optimizer}. Section \ref{sec: results} presents simulation results of a 2-speed electric vehicle for the effectiveness of the proposed strategy. Section \ref{sec: conclusion} provides conclusions and prospects.
\section{Modelling and Problem Formulation}\label{sec: Model}
\subsection{EV Dynamics Modelling}
In this section, the longitudinal motion of EV is simplified as one states namely longitudinal speed $v$ described as
\begin{equation}\label{eq: Dynamics Model}
    \Dot{v} = \dfrac{F_{\rm t} + F_{\rm b} - F_{\rm w}}{\delta m} - a_{\rm f},
\end{equation}
where $F_{\rm t}$ is the traction force empowered by the driving motor, $F_{\rm b}$ is the braking force produced by mechanical actuators, $F_{\rm w}$ is the air resistance related to travelling speed in the absence of wind, $m$ is the vehicle mass, $\delta$ is the rotating mass conversion coefficient, $a_{\rm f} = \frac{1}{\delta} \left(g \sin{\alpha} + f g \cos{\alpha}\right)$ represents the road resistance, which is corresponding with the rolling resistance factor $f$, the gravity acceleration $g$ and the road slope $\alpha$.
\par
The traction force ($F_{\rm t}$) can produce both positive and negative force for EVs to drive or brake according to the operation mode of the motor. Generally, the mechanical braking force ($F_{\rm b}$) may also exist to guarantee safety while braking. These can be named as the driver command force
\begin{equation}
\begin{aligned}
\label{eq: driving and braking torque}
F_{\rm t}+F_{\rm b}=\dfrac{{T_{\rm m} I_{\rm g} I_{0}}\eta_{\rm t}}{r_{\rm w}}+\dfrac{T_{\rm b}}{r_{\rm w}},
\end{aligned}
\end{equation}
where $T_{\rm m}$ is the motor torque, $I_{\rm g}$ is the transmission gear ratio, $I_{\rm 0}$ is the final drive ratio, $\eta_{\rm t}$ is the transmission efficiency, $r_{\rm w}$ is the wheel radius, $T_{\rm b}$ is the mechanical braking torque.
\par
Notably, $F_{\rm w}$ leading to the nonlinear features of dynamics can be described as the quadratic polynomial of $v$
\begin{equation}
\label{eq: air resistance}
    F_{\rm w} = \dfrac{\rho C_{\rm d} A_{\rm f} v^2}{2} = A_{\rm v} v^2,
\end{equation}
where $\rho$ is the air density, $C_{\rm d}$ is the aerodynamic drag coefficient, $A_{\rm f}$ is the frontal area of the vehicle, and $A_{\rm v}$ is introduced to simplify the description.
\par
Then, combining Eqs. (\ref{eq: Dynamics Model}),(\ref{eq: driving and braking torque}),(\ref{eq: air resistance}) the system dynamics is derived as
\begin{equation}
\label{eq: Integrated Dynamics}
    \dot{v} = -\dfrac{A_{\rm v}}{\delta m}v^2 + \dfrac{T_{\rm m} I_{\rm g} I_{\rm 0} \eta_{\rm t}+T_{\rm b}}{r_{\rm w}\delta m}-\dfrac{g\sin{\alpha}+fg\cos{\alpha}}{\delta}.
\end{equation}
\subsection{Problem Formulation}
For the ecological gearshift strategy, we consider applying model predictive control to achieve an optimal integer solution. The objective is to schedule the gear position to minimize energy consumption $W$. At the same time, the driving/braking torque $T_{\rm t} :=[T_{\rm m}^\top,T_{\rm b}^\top]^\top$ should be also determined given the transmission gear. Since EVs can regenerate braking energy, $T_{\rm m}$ can be also the braking torque besides $T_{\rm b}$.
\par
During the driving task of EVs, the motor power $P_{\rm m}$ is introduced to explicitly describe the objective function. Generally, $P_{\rm m}$ is modeled as a function of the motor output torque $T_{\rm m}$, the motor speed $n_{\rm m}$ and the electric conversion efficiency $\eta_{\rm m}$. However, the energy flow is inverse during the driving and braking process of the motor, such as
\begin{equation}
    \label{eq: Motor Power}
    P_{\rm m}(n_{\rm m}, T_{\rm m})=
    \begin{cases}
        \dfrac{T_{\rm m} n_{\rm m}}{\eta_{\rm m}\left(n_{\rm m}, T_{\rm m}\right)},\quad &{\rm driving},\\
        T_{\rm m} n_{\rm m}\eta_{\rm m}\left(n_{\rm m}, T_{\rm m}\right),\quad &{\rm braking},
    \end{cases}
\end{equation}
and $\eta_{\rm m}$ is mapped with motor torque and speed according to the actual test data. Meanwhile, the motor speed is directly related to vehicle speed and transmission gear ratio, as followed
\begin{equation}
\label{eq: Motor Speed}
    n_{\rm m} = \dfrac{30 i_0}{\pi r_{\rm w}} I_{\rm g} v.
\end{equation}
For the sake of simplicity, an approximate expression is necessary and the motor power can be redescribed by a polynomial function based on characteristics of motor test data, such as
\begin{equation}
    P_{\rm m}\left(n_{\rm m}, T_{\rm m}\right) = \sum_{i = 0}^2 \sum_{j = 0}^2\rho_{i,j}T_{\rm m}^i n_{\rm m}^j,
\end{equation}
where $\rho_{i,j}$ is  the fitting constant factor. By combining with Eq. (\ref{eq: Motor Speed}), it can be derived as
\begin{equation}\label{eq: Detailed Motor Power}
    P_{\rm m}\left(v,I_{\rm g},T_{\rm m}\right) = \sum_{i = 0}^2 \sum_{j = 0}^2\phi_{i,j}T_{\rm m}^i {v}^j {I_{\rm g}}^j,
\end{equation}
here, $\phi_{i,j}$ is determined by $\rho_{i,j}$ and vehicle parameters.
We should note that the above motor power description with variables is prefixed because the reference speed and acceleration are known beforehand in this problem, which is defined as $p$. In other words, the desired fraction force can be easily obtained according to Eq. (\ref{eq: Integrated Dynamics}) and the motor torque can be easily obtained once the transmission gear is determined. Here we select  $u := I_{\rm g}$ as control input and $x := W$ as state variable with
\begin{equation}
    \label{eq: Energy consumption calculation}
    \dot{W} = P_{\rm m}(I_{\rm g},p).
\end{equation}
\begin{Remark}
    \textit{The speed and acceleration trajectory reference within the prediction horizon is considered to be completely known as the parameters in this paper. In practice, the state reference should be predicted under specific driving conditions.}
\end{Remark}
\par
Hence, we define an optimal control problem to optimize gearshift with prediction horizon $T$. Considering that the sensor sampling and actuator driving process of vehicle application are under fixed period, the mixed-integer optimal control (MIOCP) is described in the discrete time with fixed time interval $\Delta t$, such as
\begin{mini!}[1]
    {U, X}{x_N}
    {\label{OCP Objective}}{}
    \addConstraint{x_0-\hat{x}_0}{= 0}
    \addConstraint{x_{k+1}}{=f\left(x_k,u_k,p_k,\Delta t\right), \quad & k = 0, \ldots, N-1} \label{eq: cstr Discrete Dynamics}
    \addConstraint{u_k}{\in \left\{I_{\rm g,1},I_{\rm g,2}\right\}, & k=0,\ldots,N-1,}
\end{mini!}
where $k$ is the index of discrete steps, $N$ is the discrete step, $U := \left[u_0,\ldots,u_{N-1}\right]^\top$, $X: =\left[x_0,\ldots,x_{N-1}\right]^\top$ are the control input and state variable sequences of the prediction horizon, respectively, $\hat{x}_0$ is the initial state, $f(\cdot,\cdot,\cdot,\cdot)$ is the numerical integration of Eq. (\ref{eq: Energy consumption calculation}) within $\Delta t$, $x_{N}$ represents the energy consumption with prediction time and we define $P:=\left[p_0,\ldots,p_{N-1}\right]^\top$. In this paper, EV is equipped with a 2-speed transmission and $I_{\rm g,1},I_{\rm g,2}$ represents the first and second gear ratio, respectively.
\subsection{Outer Convexification}
The concept of outer convexification, as applied to MIOCPs~(\cite{sager2009reformulations}), involves reformulating the ordinary differential equation using affine binary controls $b(t) \in \{0, 1\}^{n_{b}}$, where $n_{b}$ is the number of integer variables. This reformulation aims to eliminate the constraints associated with integer controls, which achieve relaxation of $b(t) \in [0,1]^{n_{b}}$ for the general optimal control problem (OCP) solution. After that, the gradient descent based approach can be applied to the relaxed OCP.
\par
Hence, we introduce binary controls to describe the selection of a 2-speed transmission. Each $b_{i}(t)$, for $i \in [1,\ldots,n_b]$ $t \in [0,T]$ represents gear ratios of $I_{\rm g,1}$ or $I_{\rm g,2}$, which indicates $n_b = 2$.
\begin{equation}
    b_{i}(t) = 1 \Leftrightarrow u(t) = I_{{\rm g},i}. \label{eq: Outer Convexification}
\end{equation}
Here, the system dynamics is discretized as Eq. (\ref{eq: cstr Discrete Dynamics}). During the period of $[k\Delta t,(k+1)\Delta t]$, the binary controls are set as constant. By combining Eq. (\ref{eq: cstr Discrete Dynamics}),(\ref{eq: Outer Convexification}), we reformulate the system dynamics and obtain the convexified dynamics in the discrete time domain for $k = 0,\ldots,N-1$.
\begin{subequations}
\begin{align}
    x_{k+1} &=\sum_{i = 1}^{n_b} b_{i,k} f_i\left(x_{k},p_{k},\Delta t\right), \\
    1 &= \sum_{i = 1}^{n_b} b_{i,k}, \label{eq: binary constraint}
\end{align}    
\end{subequations}
where $f_i(\cdot,\cdot,\cdot)$ represent the discrete system dynamics while $I_{\rm g} = I_{{\rm g},i}$. Eq. (\ref{eq: binary constraint}) is essential as it ensures that EV operates exclusively in a specific mode, such as being in first gear or second gear.
\section{Neural Network Optimizer with Integer Constraints}\label{sec: NN optimizer}
\subsection{Neural Network Optimizer Scheme}
\begin{figure}[h]
    \centering
    \includegraphics[width = 8.4cm]{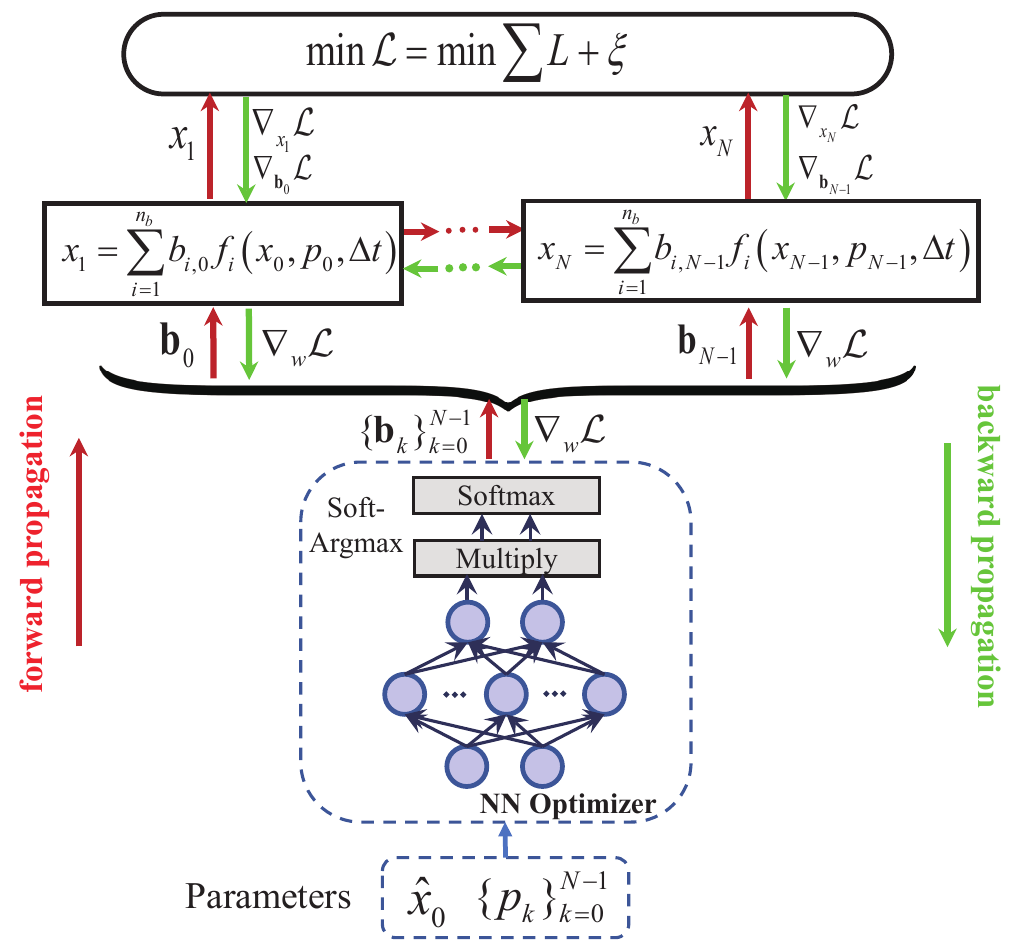}
    \caption{Neural Network Optimizer Scheme.}
    \label{fig:NN optimizer scheme}
\end{figure}
Based on the aforementioned OCP, we consider applying the MPC scheme in finite prediction horizon to achieve such ecological gearshift strategy, in which the system state can be updated according to sensor information to achieve receding control effect and the system constraint can be easily satisfied. However, it still poses the challenge of applying the algorithm to vehicle systems in real-time since MIPs are $\mathcal{NP}$-hard (\cite{russo2023learning}).
\par
In this section, we propose a neural network optimizer (NN optimizer) integrated with soft-argmax operator to help approximate the optimal integer law. As shown in Fig. \ref{fig:NN optimizer scheme}, the network will introduce the principle of MPC to achieve the prediction effect of the system state.
\par
During the training process, NN optimizer will generate all the control input sequences of the prediction horizon. Then the prediction process is considered to apply the control input to system dynamics to gain all the state information. The loss function can be reformulated according to the states and control inputs by integrating the original OCP cost function. Finally, the network will update all parameters by the gradient descent based approach.
\par
However, the solution control inputs should satisfy the meets of integer property and it's hard to enforce neural network to output the integer controls or specific types of controls unless some post-processing methods are applied. Note that amounts of post-processing methods lack the gradient information and the parameters of the network won't be updated.
\par
Inspired by the fact that softmax can handle binary classification problems, we consider applying the softmax function to achieve the solution of binary controls. Even the binary controls output by the softmax function are near 0 or 1. The control solution of MIOCP is still relaxed into $[0,1]^{n_{\rm b}}$. 
\par
\begin{figure}[h]
    \centering
    \includegraphics[width = 8cm]{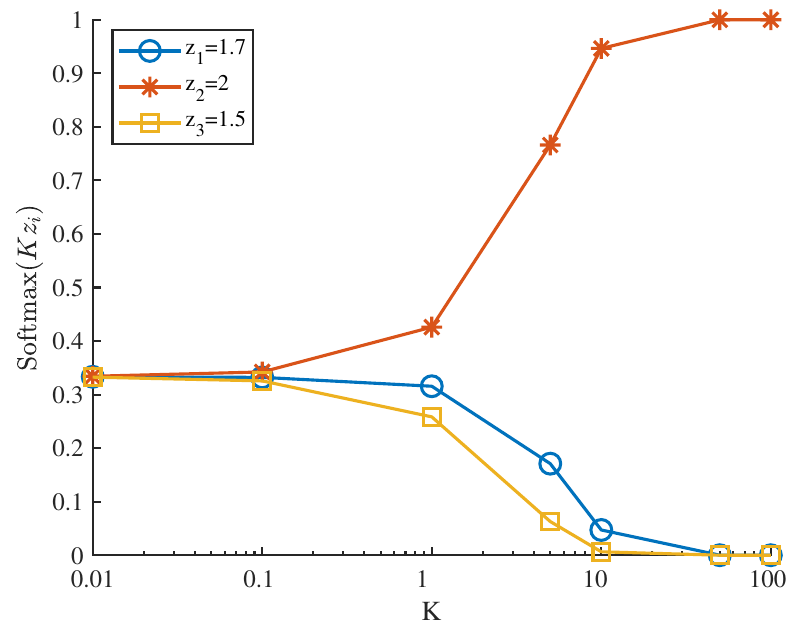}
    \caption{Soft-argmax for binary controls.}
    \label{fig:soft-argmax curves}
\end{figure}

To push the binary controls to two sides of $\{0,1\}^{n_b}$, the output layer activation with the soft-argmax operator is applied furthermore, as shown in Fig. \ref{fig:NN optimizer scheme}. The number of output nodes from each softmax layer is $n_{b}$, which can approximate the relaxed integer controls $\{0,1\}^{n_b}$ according to the probability property of the activation function. Following the soft-argmax operator, the multiplied operation should be applied to help distinguish all the elements.
\begin{equation}
    \text{Softmax}(K \cdot z_i) = \dfrac{e^{K z_i}}{\sum_j{e^{K z_j}}}.
\end{equation}
Here, $K$ is the multiple factor, $z$ is the input vector of softmax layers, and $z_i,z_j$ are both the input elements. Fig. \ref{fig:soft-argmax curves} shows that if the input elements are nearly equal ($z_1 = 1.7,z_2 = 2,z_3 = 1.5$), soft-argmax activation function ($K=1$) can help to tighten the relaxation of $\left[0,1\right]^{n_b}$. Besides, while $K\rightarrow +\infty$, the activation function will push the variables near each other towards 0 or 1. 
\subsection{Loss Function}
Then, we reformulate an integrated loss function by combining the original MIMPC cost function and outer convexification. During the training process, the amounts of training dataset should be sorted out for applicability of operating conditions. Here, we define the sample of the training dataset as the unique driving condition, which includes the control input, state, and parameter trajectory.
\par
For convenience in subsequent expressions, we define the output as $B := \left[\mathbf{b}^\top_0,\ldots,\mathbf{b}^\top_{N-1}\right]^\top$, with $\mathbf{b}_{k} := \left[b_{1,k},\ldots,b_{n_b,k}\right]^\top$. Besides, to keep only $B$ as variables, the state should be eliminated as $\overline{X}:=\left[\hat{x}_0,\ldots,\overline{x}_{N-1}\right]^\top$ with
\begin{equation}\label{eq: single shooting}
    \overline{x}_{k+1}\left(\hat{x}_0,B,P\right) = \sum_{i = 1}^{n_b}b_{i,k} f_i\left(\overline{x}_{k}\left(\hat{x}_0,B,P\right),P\right).
\end{equation}
\par
Hence, the training process of the NN optimizer can be described as the following optimization problem
\begin{equation}
    \min_{\Theta} \quad {\sum_{j=1}^{N_{\rm t}}{\omega_1 L\left(\overline{X}_j(B_j,P_j),B_j,P_j\right) + \omega_2 \xi\left(B_j\right)}, \label{eq: NN loss function}}
\end{equation}
with
\begin{align}\label{}
    F(\overline{X}_j(B_j,P_j),B_j,P_j) &= 0, \notag \\
    L\left( \overline{X}_j(B_j,P_j),B_j,P_j \right) &= \overline{x}_j(N) ,\notag    \\
    \xi \left(B_j\right) &= B_j (1 - B_j), \notag
\end{align} 
where $j$ is the index of the scenario sample, $N_t$ denotes the training data set, $w_1$, $w_2$ are weights used to trade-off between different loss terms, $F\left(\cdot,\cdot,\cdot\right)$ is the dynamic equation that consolidates all discrete states by following Eq. (\ref{eq: single shooting}), and $\Theta$ are decision variables representing weights and biases of the neural network.
\par
Even the soft-argmax can nearly get binary variables, this method can't make an effect once input elements are too close. As a result, we introduce the penalty term $\xi\left(B_j\right)$ for control input $B_j$ to obtain the solution of $\{0,1\}^{n_b}$.
\subsection{Parameters Updating}
Since the forward pass process has ended, NN optimizer parameters should be updated according to the gradient descent method. After iterations, the training loss will converge to a near-minimum. A general backward updating method of network parameters can be described as
\begin{equation}
    \Theta^{(n+1)} = \Theta^{(n)} - \eta \nabla_{\Theta^{(n)}}\mathcal{L},
\end{equation}
and $\mathcal{L}$ is the defined network loss function of Eq. (\ref{eq: NN loss function}), the superscript $n$ denotes the iteration epoch, $\eta$ represents the learning rate of training process, and $\nabla_{\Theta^{(n)}}\mathcal{L}$ describes the gradient of loss function with respect to the network parameters.
\par
Then, the control binary output will be updated in the training process (\cite{bottcher2022ai})
\begin{align}
    \mathbf{b}_k(\Theta^{(n+1)}) &= \mathbf{b}_k(\Theta^{(n)}+\Delta \Theta^{(n)}) ,\notag \\
    &=\mathbf{b}_k(\Theta^{(n)}) + \mathcal{J}_{\mathbf{b}_k}\Delta \Theta^{(n)},
\end{align}
\par
where $\mathcal{J}_{\mathbf{b}_k}$ is the Jacobian of $\mathbf{b}_k$ with elements $(\mathcal{J}_{\mathbf{b}_k})_{ij} = \partial b_{i,k}/\partial \Theta_j$, and $\Delta \Theta^{(n)} = -\eta\nabla_{\Theta^{(n)}}\mathcal{L}$.
\par
Notably, all the training process is based on gradient and our proposed simplified and relaxed approach will avoid that ``gradient does not exist". And according to the chain rule, the updating principle is extensively denoted as
\begin{align}
    \mathbf{b}_k(\Theta^{(n+1)}) &= \mathbf{b}_k(\Theta^{(n)}) - \eta \mathcal{J}_{\mathbf{b}_k}\mathcal{J}_{\mathbf{b}_k}^\top\nabla_{\mathbf{b}_k}\mathcal{L},
\end{align}
and
\begin{align}
\nabla_{\mathbf{b}_k}\mathcal{L} = \dfrac{\partial \overline{X}}{\partial \mathbf{b}_k}\nabla_{\overline{X}}\mathcal{L}_1 + \nabla_{\mathbf{b}_k}\mathcal{L}_2,
\end{align}
where $\mathcal{L}_1$ and $\mathcal{L}_2$ represent first and second term of Eq. (\ref{eq: NN loss function}), respectively. Here, $\mathcal{L}_1$ is related to control input, state, and parameter trajectory. $\mathcal{L}_2$ is only related to binary output $\mathbf{b}_k$ from neural network.
\subsection{Closed-loop Application}
\begin{figure}[h]
    \centering
    \includegraphics[width = 8.4cm]{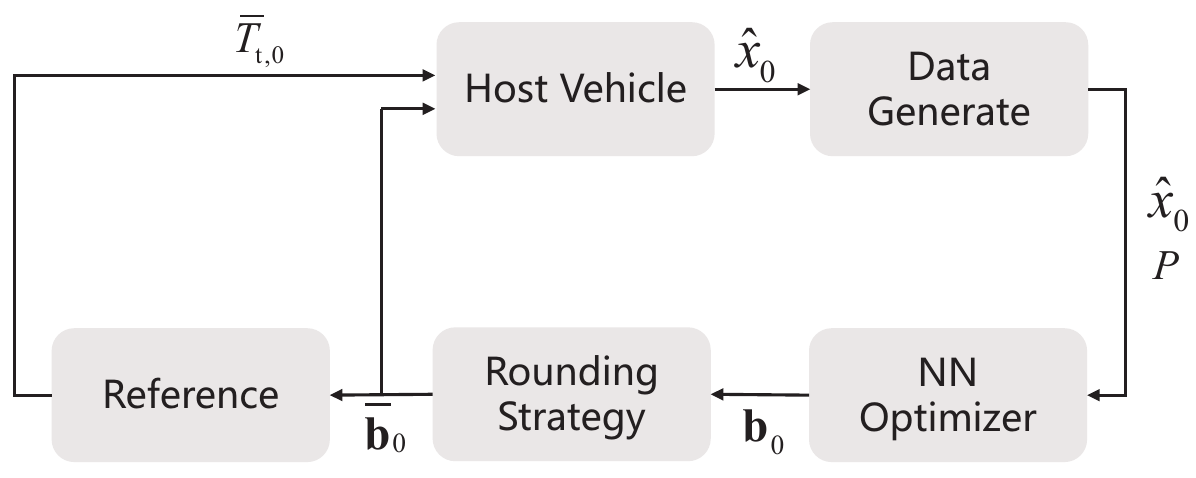}
    \caption{Online control strategy.}
    \label{fig: Closed Loop application}
\end{figure}

While deploying the trained NN optimizer to the ecological gearshift strategy, some necessary operations should be introduced to achieve the effect of closed-loop control and enforce the network output integer variables $\{0,1\}^{n_b}$. Fig. \ref{fig: Closed Loop application} shows the overall online control strategy for the gearshift.
\par
 Following MPC, the first element $\mathbf{b}_0$ of binary control input sequence $B = \left[\mathbf{b}^\top_0,\ldots,\mathbf{b}^\top_{N-1}\right]^\top$ is applied during the closed control process. And the driving/braking torque $T_{\rm t,0}$ at present will also be got accordingly.
 \begin{Remark}
     \textit{Please note that NN optimizer can achieve binary solutions in the majority of cases. However, occasional instances may result in solutions very close to integers (e.g., 0.997 or 0.003) due to numerical accuracy considerations. It's necessary to apply a standard rounding strategy (\cite{sager2005numerical}) to force these outputs into integers during actual application.}
 \end{Remark}
\section{Results and Analysis}\label{sec: results}
This section presents the numerical simulations of the gearshift strategy to effectively verify our designed control architecture. The neural network is trained under \texttt{PyTorch} and transferred into \texttt{Matlab} for real-time application. Most simulations were conducted on a personal computer with Intel$\circledR$ Core$^\text{TM}$ i7-12700H @ 2.70GHz and 16GB RAM.
\par
We perform it on the New European Driving Cycle (NEDC) for EV with a 2-speed transmission, which could represent the normal driving condition of the real world and poses a challenge to mixed integer programming due to the long-time domain and frequently changing working conditions. Besides, the vehicle parameters are listed in Table \ref{tab:Vehicle Parameters}. Here, the prediction horizon $T$ is set as 8 s and the sampling time both as simulation interval $\Delta t$ is set as 1 s for direct acquisition of the accurate driving cycle data.
\begin{table}[h]
    \caption{Main parameters of electric vehicle.}
    \centering
    \renewcommand\arraystretch{1}
    \begin{tabular}{cccc}
    \toprule
         Parameter  &Value   &Parameter   &Value\\
    \midrule
         $m$       &1533kg    &$I_{\rm g}$     &[3.4 1.5]\\
         $f$       &0.01   &$I_0$    &3.94\\
         $\eta_{\rm t}$  &0.96 &$A_{\rm v}$    &0.45864kg/m\\
    \bottomrule
    \end{tabular}
    \label{tab:Vehicle Parameters}
\end{table}
\par
The architecture utilized in our network consists of fully connected layers, including 2 hidden layers with 64 neurons. Each hidden layer is activated by the hyperbolic tangent function. As for the output layer, we set $8$ pairs of softmax layers with $2$ neurons. Meanwhile, the number of pairs represents the gearshift decision sequence with the same as $N$ and the number of neurons represents the binary control input $n_b$. We select the multiplied math operation $K = 10$ to approximate integer output and the learning rate is $0.001$.
\par
The training data is set as the driving cycle data with sliding windows. Each window includes the initial state $x_0$, the reference speed and acceleration $P$. The sliding window moves from the beginning to the end of the driving cycle. After 300 epochs, the loss of NN optimizer converged.
\par
To show the effectiveness of the solution, the nonlinear optimization tool \texttt{CasADi}~(\cite{Andersson2019}) with MIP solver \texttt{Bonmin}~(\cite{bonami2007bonmin}) is compared to solve the same MIMPC problem. Besides, we set the rule-based method as the baseline to verify the energy-saving effect.
\begin{figure}[h]
    \centering
    \includegraphics[width=8cm]{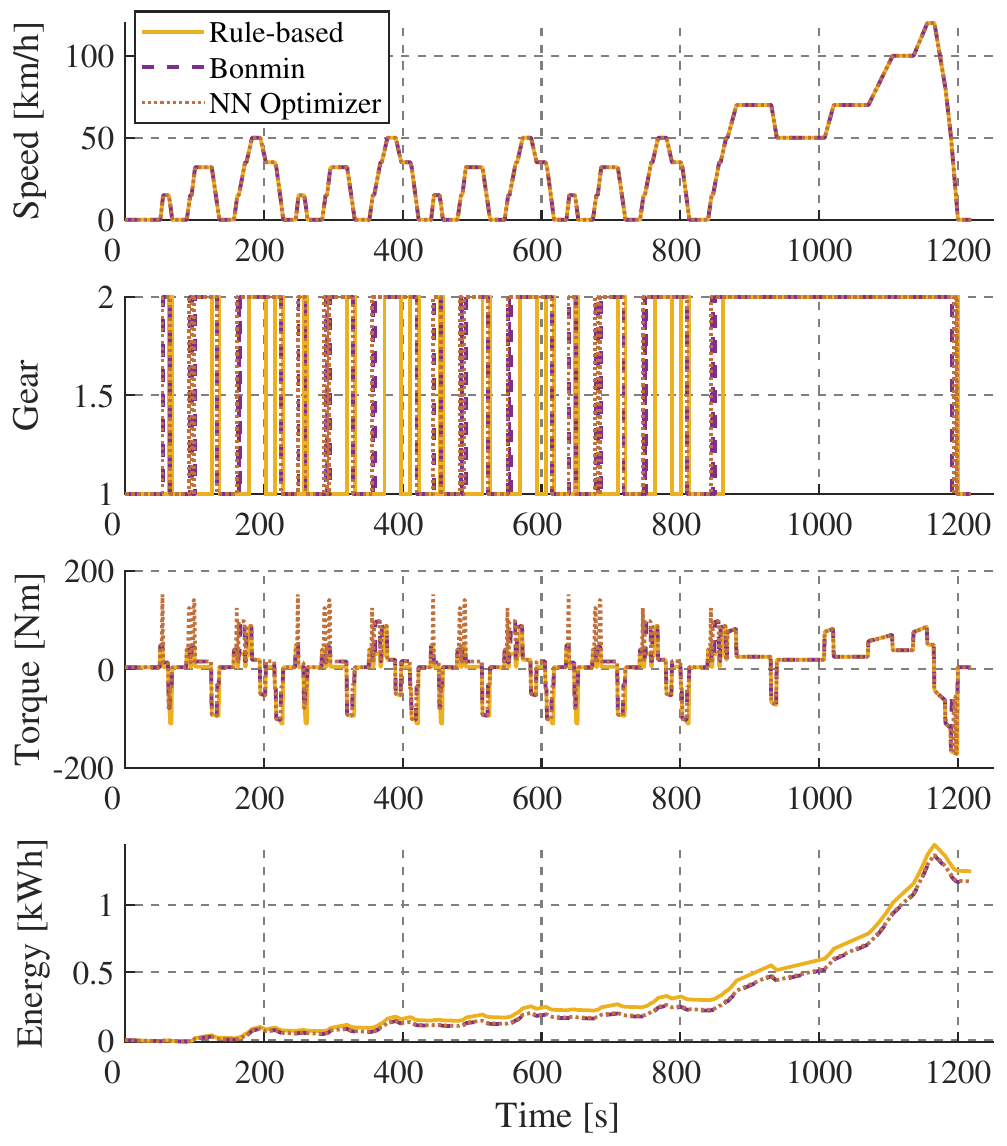}
    \caption{Gearshift comparison for NEDC of 2-speed EV.}
    \label{fig:Sim_NEDC}
\end{figure}
\par
In Fig. \ref{fig:Sim_NEDC}, the algorithms can help to ecologically gear shift. The proposed NN optimizer with soft-argmax operator is capable of obtaining integer solutions that are close to those achieved by \texttt{Bonmin}, except for a few operating conditions, such as $300s, 500s$, and $700s$. It is precisely in these operating conditions that our algorithm exhibits a gap in energy consumption performance due to there existing peak points for torque.  
\par
Table. \ref{tab: Performance comparison} shows the computation time and energy consumption of each gearshift strategy. The NN optimizer can save energy by 6.02\% compared rule-based strategy and achieve 0.55\% sub-optimality of the mature MIP solver \texttt{Bonmin}. In terms of the computation time, We deployed the NN optimizer on the dSPACE MicroAutoBox-III DS1403 rapid prototyping units with four ARM Cortex-A15 processor cores. Meanwhile, NN optimizer consumes 0.045ms and substantially improves the algorithm's solving speed, in which \texttt{Bonmin} can't finish optimization in real-time within 1s sampling period. Hence, we conducted tests on the computation time of \texttt{Bonmin} on a personal computer, and it averaged a consumption of 229ms per solution.
\begin{table*}[t]
    \caption{Performance comparison.}
    \centering
    \label{tab: Performance comparison}
    \begin{tabular}{ccccc}
    \toprule
    \multirow{2}{*}{Method} & \multirow{2}{*}{Energy} & \multirow{2}{*}{Rate} & \multicolumn{2}{c}{Computation time}     \\\cline{4-5}
                       & &                   & \multicolumn{1}{c}{mean} & worst \\ 
        \midrule
        Rule-based  &1.2470kWh  &-  &-  &-\\
            \texttt{Bonmin}      &1.1651kWh   &6.57\%  &229ms [CPU]   &2315ms [CPU]\\
        NN optimizer &1.1720kWh     &6.02\%   &0.045ms [dSPACE]  &0.053ms [dSPACE]\\
    \bottomrule
    \end{tabular}
    \label{tab:my_label}
\end{table*}
\par
\begin{figure}[h]
    \centering
    \includegraphics[width=8cm]{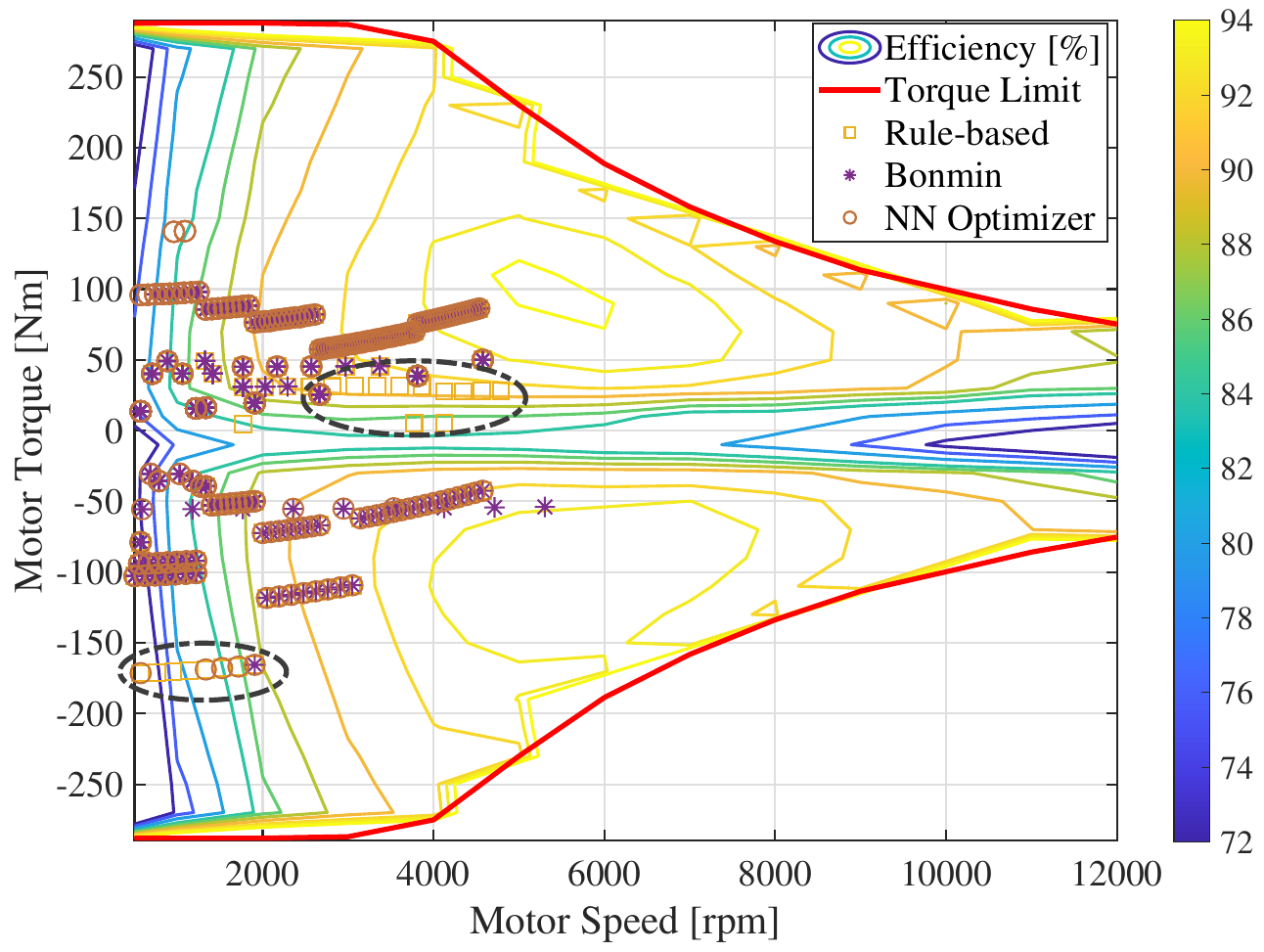}
    \caption{Motor efficiency comparison of working condition.}
    \label{fig: Motor Efficiency}
\end{figure}

To analyze the energy-saving effect, we labelled the motor working condition of the three algorithms in the motor efficiency map as shown in Fig. \ref{fig: Motor Efficiency}. In most driving conditions, rule-based, \texttt{Bonmin}, and NN optimizer can get the same working points. It is precisely in the operating conditions represented by the blue dashed line in the figure that the efficiency of the motor decreases, leading to an increase in energy consumption. The higher the frequency of occurrence of these points, the greater the difference in energy consumption.
\section{Conclusion}\label{sec: conclusion}
In this paper, we propose an online gearshift strategy for vehicles equipped with transmission to achieve eco-driving. By introducing the outer convexification method, the binary controls in the MIMPC problem are relaxed to be continuous. In addition, we employ neural networks to achieve the binary controls of MIMPC, in which soft-argmax operator is integrated into the forward updating process to approximate the integer solution. Finally, the receding solving method and standard rounding strategy are applied for closed-loop control. The numerical simulation results indicate that the method can help to gearshift ecologically and reduce the calculation burden significantly.
\par
In the future, we will further apply the scheme to more general MIMPC problems with both continuous and integer controls, in which the state variable constraints and combinatorial constraints should also be considered.

\bibliography{ifacconf}             
\end{document}